\documentclass[conference,a4paper,10pt]{IEEEtran}

\usepackage[T1]{fontenc}
\usepackage{amsmath,amsthm,amssymb,stackrel}
\usepackage{mathtools}
\usepackage[utf8]{inputenc}
\usepackage{pgfplots,tikz}
\usepackage{subfigure}
\usepackage{xcolor}
\usepackage{xspace}
\usepackage{hhline}
\usepackage{csquotes}
\usepackage{hyperref}
\usepackage{cuted}
\usepackage{flushend}
\usetikzlibrary{patterns}
\usepgfplotslibrary{external} 

\newcommand{\figref}[1]{Fig. \ref{#1}}

\newcommand{\N}{\ensuremath{\text{N}}\xspace}

\newcommand{\KP}{\ensuremath{\text{K}_\text{P}}\xspace}
\newcommand{\NP}{\ensuremath{\text{N}_\text{P}}\xspace}

\newcommand{\eqdef}{\stackrel{\text{def}}{=}}

\theoremstyle{definition}

\definecolor{mygreen}{rgb}{0, 0.75, 0}

\begin{document}
	\title{Optimized pilot distribution to track the phase noise in DFT-s-OFDM for sub-THz systems}
	\author{
		\IEEEauthorblockN{
			J.-C. Sibel, V. Corlay, and A. Bechihi
		}
		\IEEEauthorblockA{
			Mitsubishi Electric R\&D Centre Europe\\
			Rennes, France\\
			Email: \{j.sibel, v.corlay, a.bechihi\}@fr.merce.mee.com
		}
	}
	\maketitle

\begin{abstract}
	In this paper, we focus on the selection of a pilot pattern to track the phase noise in high frequency bands in a DFT-s-OFDM chain. By considering the Wiener filter at the receiver side to perform the tracking, we use the inner cost function of the said filter as the cost function for the pilot selection. To obtain this cost function, the phase noise autocorrelation is required. Therefore, we introduce a new mathematical approximation of the autocorrelation function of the practical 3GPP phase noise model. At first, this leads to an analytical expression of the Wiener filter coefficients. Then, the said coefficients allow us to obtain an analytical expression of the cost function. Thus, by means of this result, we are able to provide a pilot pattern that  jointly satisfies a constraint on the pilot overhead and a constraint on the minimum performance of the Wiener filter.
\end{abstract}


\section*{Introduction\label{sec:Introduction}}
	3GPP NR specifications provide the 5G technology to take benefit of high frequency ranges, namely mmWaves, until 71GHz \cite{3GPP38.101-2}. It is expected that sub-THz frequencies and beyond will be used for the 6G technology \cite{shafie2022terahertz}. This opens the room for greater bandwidths, higher data rates, etc. However, working with such high frequency values is not straightforward because of the limited capability of hardware materials. Among others, the phase noise is a phenomenon that stems from this limitations and that induces an increasing and negative impact on signals as the carrier frequency increases. Even though some futuristic components might be developed to cancel this limitation, the 3GPP specifications provide dedicated pilots called Phase Tracking Reference Signals (PT-RS) that enable to estimate and track the phase noise to work with good communication performance \cite{3GPP38.211}.
	
	The PT-RS can be distributed in several manners depending on the waveform, the bandwidth, the subcarrier spacing, the Modulation and Coding Scheme (MCS), etc. In this paper, we consider an Orthogonal Frequency-Division Multiplexing (OFDM) chain with a Discrete Fourier Transform (DFT) precoding, simply called DFT-s-OFDM \cite{DFTsOFDM}. For this case, the PT-RS are inserted in the time domain before the DFT precoding in equally spaced groups. During the phase of specifications, the decisions on the values of group sizes and group spacings were made based on numerical evaluations of the whole DFT-s-OFDM chain from several companies \cite{3GPPRAN1AH02}. The drawback in this approach is that results also depends, at least, on the MCS that is out of the tracking algorithm. It would be more relevant to select the PT-RS pattern based on the performance of the tracking algorithm only.
	
	
	In this paper, we focus on the Wiener filter \cite{Oppenheim2010} to track the phase noise which is a usual scheme used in wireless communications. The filter coefficients are obtained by minimizing a cost function $J$ that is determined by the autocorrelation function of the phase noise. The analytical expression is not available for the practical model of 3GPP \cite{3GPP38.808} usually used in the literature. To overcome this difficulty, we propose an analytical approximation of the said autocorrelation function based on its graphical shape which helps obtain an analytical expression of $J$. Out of this effort, we are given the possibility to analyze and predict the behavior of $J$ as a function of the simulation parameters, e.g., the carrier frequency and the PT-RS spacing. Indeed, including one constraint related to an arbitrary maximum value of $J$ and one constraint related to a maximum overhead for the sake of the spectral efficiency, we are able to extract the PT-RS spacing that minimizes $J$, i.e., that offers the best performance of the Wiener filter. As in the 3GPP specifications, we consider the PT-RS to be equally spaced and we simplify the study by assuming that the PT-RS groups are of size one. 
	
	\textbf{Contributions -} We provide an analytical model of the autocorrelation function for the 3GPP phase noise model. Based on this, we provide the analytical expression of the Wiener filter coefficients. From this, we derive an analytical form of the cost function $J$ as a function of the PT-RS spacing. We show that through numerical evaluations a linear approximation of $J$ is equivalent. We finally show how to select the PT-RS spacing that satisfies a constraint on $J$ as well as a constraint on the PT-RS overhead.

	The paper is structured as follows. Section \ref{sec:BasicSetting} presents the considered communication chain and describes our approximation of the 3GPP phase noise model. Then, Section \ref{sec:WienerFilter} exposes the Wiener filter computation with the derivation of the filter coefficients based on our approximation model. Section \ref{sec:CostFunctionDerivation} presents the associated derivation of the cost function of the Wiener filter using the previously computed filter coefficients and a analysis of its behavior. Finally, Section \ref{sec:PTRSExtraction} describes the method for obtaining the PT-RS spacing that fulfills a maximum cost constraint and a maximum overhead constraint with an application example. 
	
\section{Phase noise in DFT-s-OFDM\label{sec:BasicSetting}}
	This section presents the DFT-s-OFDM chain and an analytical approximation of the 3GPP phase noise model. 
	\subsection{Communication chain\label{sec:CommunicationChain}}
		We consider a single DFT-s-OFDM symbol $\underline{x} = [x_1 \dots x_{\N}]$ of length $N$ where $x_n$ is a constellation symbol (QAM, PSK). The phase noise $\underline{\alpha} = [\alpha_1 \dots \alpha_{\N}]$ affects the signal, where $\alpha_n = e^{i\phi_n}$. Then, an AWGN $\underline{z} = [z_1 \dots z_{\N}]$ also alters the signal,  where $z_n \sim \mathcal{CN}(0,\sigma_z^2)$. Hence, we get the received signal $\underline{y} = [y_1 \dots y_{\N}]$ with:
		\begin{equation}
			y_n = \alpha_n x_n + z_n.
		\end{equation}
		We allocate $\NP$ pilots whose indexes in the symbol $\underline{x}$ are $p_1, \dots, p_{\NP}$. We assume the pilots to be uniformly distributed such that the position difference between any two consecutive pilots $p_j,p_{j+1}$ is $\Delta = |p_{j+1}-p_j|$. Accordingly, for $1\leq j\leq \NP$, the $j^{\text{th}}$ pilot index is:
		\begin{align}
			p_j = p_1 + (j-1)\Delta.
			\label{eq:pj}
		\end{align}
		On each pilot index $j$, provided that $|x_{p_j}|^2=1$, we assume a least-square (LS) estimate of the phase noise:
		\begin{equation}
			\tilde{\alpha}_{p_j} \eqdef y_{p_j}\bar{x}_{p_j},
		\end{equation}
		where $\bar{x}_{p_j}$ stems for the complex conjugate of $x_{p_j}$. This amounts to:
		\begin{equation}
			\tilde{\alpha}_{p_j} = \alpha_{p_j} + \tilde{z}_{p_j},
		\end{equation}
		where $\tilde{z}_{p_j} = z_{p_j}\bar{x}_{p_j}$ follows the same law as $z_{p_j}$. The performance of any phase noise tracking process become relevant when working at high Signal-to-Noise Ratio (SNR) values, see the evaluation results in \cite{Sibel2022}. Consequently, we assume in the rest of the paper that the SNR is infinite such that $\sigma_z = 0$ meaning that:
		\begin{equation}
			y_n = \alpha_n x_n.
		\end{equation}
	
	\subsection{Phase noise exponential model\label{sec:EXPModel}}
		We consider the phase noise model from the 3GPP specifications \cite[4.2.3.1]{3GPP38.808} that scales with the carrier frequency $F_c$. 
		\begin{figure}[!h]
			\centering
			\input{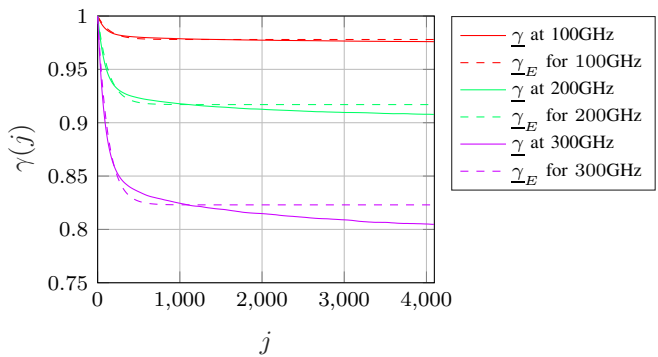}
			\caption{Empirical autocorrelation of the 3GPP model in solid lines, exponential approximation $\mathcal{E}_\N(a,b)$ in dashed lines.}
			\label{fig:MapTR3GPPToExp}
		\end{figure}
		
\begin{figure}[!h]
			\centering
			\subfigure[$a$ against $F_c$.]{\hspace*{-1em}
%
%
\begin{tikzpicture}

\begin{axis}[%
	width=0.35\linewidth,
	height=0.12\textheight,
	font=\footnotesize,
	scale only axis,
	scaled ticks=false, 
	tick label style={/pgf/number format/fixed},
	xmin=100e9,
	xmax=300e9,
	xlabel style={font=\color{white!15!black}},
	xlabel={$F_c$ (GHz)},
	xtick={100e9,150e9,200e9,250e9,300e9},
	xticklabels={$100$,$150$,$200$,$250$,$300$},
	ymin=0.0073,
	ymax=0.0079,
	yminorticks=true,
	ylabel style={font=\color{white!15!black}, anchor=south west, at={(0.2,0.18)}},
	ylabel={$1000\times a$},
	ytick={0.0073,0.0074,...,0.0079},
	yticklabels={7.3,7.4,7.5,7.6,7.7,7.8},
	axis background/.style={fill=white},
	title style={font=\bfseries},
	xmajorgrids,
	ymajorgrids,
	yminorgrids]
\addplot [color=black]
  table[row sep=crcr]{%
100000000000	0.0073584211807481\\
101000000000	0.00735700933475423\\
102000000000	0.00735461779240096\\
103000000000	0.00735369780540538\\
104000000000	0.00735192362124607\\
105000000000	0.00735109697457412\\
106000000000	0.00734978154321954\\
107000000000	0.00734951665134339\\
108000000000	0.00734864987075704\\
109000000000	0.00734870799078151\\
110000000000	0.00734830725557471\\
111000000000	0.00734758820370939\\
112000000000	0.00734760997244678\\
113000000000	0.00734736009171719\\
114000000000	0.00734655596118966\\
115000000000	0.00734629281101884\\
116000000000	0.00734570073718081\\
117000000000	0.00734578316059398\\
118000000000	0.00734542403361947\\
119000000000	0.00734571877396997\\
120000000000	0.00734567643877387\\
121000000000	0.00734543649826435\\
122000000000	0.00734579748170834\\
123000000000	0.00734602878277338\\
124000000000	0.00734589821248345\\
125000000000	0.00734629967073907\\
126000000000	0.007346570306993\\
127000000000	0.00734756637679294\\
128000000000	0.00734837184877974\\
129000000000	0.00734899679907282\\
130000000000	0.00734939583583772\\
131000000000	0.00734982210499542\\
132000000000	0.00735109618148117\\
133000000000	0.00735153967929484\\
134000000000	0.00735348240910035\\
135000000000	0.00735451059715667\\
136000000000	0.00735695788986233\\
137000000000	0.00735857290215397\\
138000000000	0.00736070914519419\\
139000000000	0.00736271763570441\\
140000000000	0.00736522687666729\\
141000000000	0.00736687624392525\\
142000000000	0.00736769001422706\\
143000000000	0.00736982126242338\\
144000000000	0.00737115078267763\\
145000000000	0.00737290275875275\\
146000000000	0.00737452515708595\\
147000000000	0.00737660839813807\\
148000000000	0.00737786920716561\\
149000000000	0.00738019210978545\\
150000000000	0.00738175092469857\\
151000000000	0.00738375137303143\\
152000000000	0.0073849797479148\\
153000000000	0.00738661151621358\\
154000000000	0.00738814336184309\\
155000000000	0.00739007064115573\\
156000000000	0.00739130331385851\\
157000000000	0.00739356727112069\\
158000000000	0.00739613903163168\\
159000000000	0.00739801687822913\\
160000000000	0.00740018280862927\\
161000000000	0.00740231087372079\\
162000000000	0.00740485320692805\\
163000000000	0.00740684524514106\\
164000000000	0.00740920491058182\\
165000000000	0.00741090610770546\\
166000000000	0.00741352571913926\\
167000000000	0.00741561729024724\\
168000000000	0.0074180326205255\\
169000000000	0.00741984535773667\\
170000000000	0.00742188711294392\\
171000000000	0.00742468448424335\\
172000000000	0.00742685418352495\\
173000000000	0.00742927503521319\\
174000000000	0.00743180504897407\\
175000000000	0.00743415244039427\\
176000000000	0.00743667683977429\\
177000000000	0.00743862125541726\\
178000000000	0.00744074306831056\\
179000000000	0.00744223181832082\\
180000000000	0.00744386876320582\\
181000000000	0.00744613405053068\\
182000000000	0.00744783711432967\\
183000000000	0.00744979108004739\\
184000000000	0.00745187273826295\\
185000000000	0.00745334338339316\\
186000000000	0.00745496270589762\\
187000000000	0.0074565661714288\\
188000000000	0.0074579109143575\\
189000000000	0.00745980981968359\\
190000000000	0.00746181366314944\\
191000000000	0.00746331610500301\\
192000000000	0.00746497543099909\\
193000000000	0.00746677135215157\\
194000000000	0.00746853772869171\\
195000000000	0.00747050502760754\\
196000000000	0.00747208979740328\\
197000000000	0.00747393265712687\\
198000000000	0.00747595485794996\\
199000000000	0.00747753869959805\\
200000000000	0.00747931188151121\\
201000000000	0.00748172206047611\\
202000000000	0.00748385596261702\\
203000000000	0.0074855786939134\\
204000000000	0.00748789848356839\\
205000000000	0.00749035110543753\\
206000000000	0.00749242078940908\\
207000000000	0.0074947014818073\\
208000000000	0.00749711432160046\\
209000000000	0.00749913913779993\\
210000000000	0.00750132838456835\\
211000000000	0.00750410635544388\\
212000000000	0.00750663690589106\\
213000000000	0.00750927586987859\\
214000000000	0.00751192399970665\\
215000000000	0.00751473225991118\\
216000000000	0.00751770759021988\\
217000000000	0.00752039290965679\\
218000000000	0.00752327580301359\\
219000000000	0.00752631921468971\\
220000000000	0.00752947727564227\\
221000000000	0.00753235751445447\\
222000000000	0.00753542178872402\\
223000000000	0.00753869783721291\\
224000000000	0.00754211068627642\\
225000000000	0.00754559036863251\\
226000000000	0.0075487294685528\\
227000000000	0.00755239842585734\\
228000000000	0.00755586207594017\\
229000000000	0.00755976035496651\\
230000000000	0.00756339633993155\\
231000000000	0.00756706242184603\\
232000000000	0.0075705399769743\\
233000000000	0.00757415505023653\\
234000000000	0.00757781629825775\\
235000000000	0.00758144936532894\\
236000000000	0.00758525269758732\\
237000000000	0.00758918277449358\\
238000000000	0.00759285371491443\\
239000000000	0.00759687250509893\\
240000000000	0.00760058072188566\\
241000000000	0.0076042919619639\\
242000000000	0.00760775117584309\\
243000000000	0.00761127446760374\\
244000000000	0.00761515025601611\\
245000000000	0.00761904897123471\\
246000000000	0.00762299907671674\\
247000000000	0.00762664234389165\\
248000000000	0.00763039584544044\\
249000000000	0.00763410533579859\\
250000000000	0.00763753597053037\\
251000000000	0.00764134618683589\\
252000000000	0.00764490632346973\\
253000000000	0.00764853863765778\\
254000000000	0.00765186526728679\\
255000000000	0.00765522766376123\\
256000000000	0.00765864974007964\\
257000000000	0.00766243498022977\\
258000000000	0.00766599931305181\\
259000000000	0.00766988910459335\\
260000000000	0.0076735295010693\\
261000000000	0.00767693719815435\\
262000000000	0.00768074003808985\\
263000000000	0.00768460352310791\\
264000000000	0.00768821132780416\\
265000000000	0.00769186624792859\\
266000000000	0.00769560968750404\\
267000000000	0.00769942785833648\\
268000000000	0.00770302060407947\\
269000000000	0.00770688333664292\\
270000000000	0.0077104700614142\\
271000000000	0.00771409342227424\\
272000000000	0.00771782578987831\\
273000000000	0.00772159407020304\\
274000000000	0.00772508964633959\\
275000000000	0.00772860337284904\\
276000000000	0.00773216171835684\\
277000000000	0.00773605909931426\\
278000000000	0.00773975348183944\\
279000000000	0.00774346013406133\\
280000000000	0.00774719153717559\\
281000000000	0.00775064042033173\\
282000000000	0.00775411060328934\\
283000000000	0.00775745548094007\\
284000000000	0.00776084463835836\\
285000000000	0.00776428242299549\\
286000000000	0.00776758876010984\\
287000000000	0.00777056572659509\\
288000000000	0.0077738382130875\\
289000000000	0.0077767930353152\\
290000000000	0.00777998440821698\\
291000000000	0.00778272416740089\\
292000000000	0.00778573854802624\\
293000000000	0.00778855312024462\\
294000000000	0.0077913699773129\\
295000000000	0.00779392090815534\\
296000000000	0.00779632785890236\\
297000000000	0.00779866820117242\\
298000000000	0.00780128223604507\\
299000000000	0.00780352329934077\\
300000000000	0.00780600324117115\\
301000000000	0.00780793850587906\\
302000000000	0.00781020580555456\\
303000000000	0.00781224667879453\\
304000000000	0.00781397240728761\\
305000000000	0.00781577083423361\\
306000000000	0.00781740084123615\\
307000000000	0.00781897040274036\\
308000000000	0.00782095642456409\\
309000000000	0.00782244909366693\\
};
\end{axis}
\end{tikzpicture}%
\label{fig:ParametersaAgainstCarrierFrequency}}
			\subfigure[$b$ against $F_c$.]{\hspace*{-1em}
%
%
\begin{tikzpicture}

\begin{axis}[%
	width=0.35\linewidth,
	height=0.12\textheight,
	font=\footnotesize,
	scale only axis,
	scaled ticks=false, 
	tick label style={/pgf/number format/fixed},
	xmin=100e9,
	xmax=300e9,
	xlabel style={font=\color{white!15!black}},
	xlabel={$F_c$ (GHz)},
	xtick={100e9,150e9,200e9,250e9,300e9},
	xticklabels={$100$,$150$,$200$,$250$,$300$},
	ymin=0.8,
	ymax=0.98,
	yminorticks=true,
	ylabel style={font=\color{white!15!black}, anchor=south west, at={(0.2,0.38)}},
	ylabel={$b$},
	axis background/.style={fill=white},
	title style={font=\bfseries},
	xmajorgrids,
	ymajorgrids,
	yminorgrids]
\addplot [color=black]
  table[row sep=crcr]{%
100000000000	0.977392014811171\\
101000000000	0.977174612199972\\
102000000000	0.976933283460017\\
103000000000	0.976657469963949\\
104000000000	0.976354166195315\\
105000000000	0.976017538743875\\
106000000000	0.975656937649559\\
107000000000	0.97527027900953\\
108000000000	0.974868275003811\\
109000000000	0.974449256570051\\
110000000000	0.974022196134097\\
111000000000	0.973570494689041\\
112000000000	0.973120123909524\\
113000000000	0.972659658943235\\
114000000000	0.972196390137711\\
115000000000	0.971720940579193\\
116000000000	0.971242647943407\\
117000000000	0.970753742275198\\
118000000000	0.970256460182309\\
119000000000	0.969753742275197\\
120000000000	0.969242647943407\\
121000000000	0.968725379806012\\
122000000000	0.968196390137711\\
123000000000	0.967664098170054\\
124000000000	0.967120123909524\\
125000000000	0.966574933915861\\
126000000000	0.966022196134097\\
127000000000	0.965473897579153\\
128000000000	0.964924268358671\\
129000000000	0.96438473332496\\
130000000000	0.963848002130484\\
131000000000	0.963315413287691\\
132000000000	0.962788070543542\\
133000000000	0.962260760951472\\
134000000000	0.961725379806012\\
135000000000	0.961187155583285\\
136000000000	0.960634209128529\\
137000000000	0.960073648432274\\
138000000000	0.95949878706864\\
139000000000	0.958914439020889\\
140000000000	0.958321484503745\\
141000000000	0.957719308589958\\
142000000000	0.95712071869288\\
143000000000	0.956520155384849\\
144000000000	0.955920669184179\\
145000000000	0.955324801738019\\
146000000000	0.954728468983815\\
147000000000	0.954132601537655\\
148000000000	0.953533115336984\\
149000000000	0.952924513627642\\
150000000000	0.9523078107225\\
151000000000	0.951688223395318\\
152000000000	0.95105802795535\\
153000000000	0.950419383700592\\
154000000000	0.949777605321065\\
155000000000	0.949124563136344\\
156000000000	0.948470305218489\\
157000000000	0.947808081024856\\
158000000000	0.947142089482143\\
159000000000	0.946478336805973\\
160000000000	0.945817692691083\\
161000000000	0.945160464669605\\
162000000000	0.944503100100667\\
163000000000	0.943853752168286\\
164000000000	0.943203957780367\\
165000000000	0.942546571446676\\
166000000000	0.941888010724623\\
167000000000	0.941218166617356\\
168000000000	0.940538533045364\\
169000000000	0.939855688879568\\
170000000000	0.939166496647625\\
171000000000	0.938470298404661\\
172000000000	0.937771573343045\\
173000000000	0.937069293992121\\
174000000000	0.936366546156327\\
175000000000	0.935660078597687\\
176000000000	0.934943976684724\\
177000000000	0.934226205018667\\
178000000000	0.933500680093328\\
179000000000	0.932764872421099\\
180000000000	0.932023036186425\\
181000000000	0.931283602973128\\
182000000000	0.93054405766875\\
183000000000	0.9298031424608\\
184000000000	0.929063876283741\\
185000000000	0.928323515986091\\
186000000000	0.9275838123695\\
187000000000	0.926841000191629\\
188000000000	0.926091876372744\\
189000000000	0.925331673768367\\
190000000000	0.924562100186943\\
191000000000	0.923789279586581\\
192000000000	0.923009223947522\\
193000000000	0.92222279493715\\
194000000000	0.921428479825785\\
195000000000	0.920638108576507\\
196000000000	0.919848232691962\\
197000000000	0.91905736445477\\
198000000000	0.9182692704144\\
199000000000	0.917482776032792\\
200000000000	0.916695791037974\\
201000000000	0.915910825138903\\
202000000000	0.915124678068749\\
203000000000	0.914334396493822\\
204000000000	0.913537554563804\\
205000000000	0.912732437177442\\
206000000000	0.911915684763598\\
207000000000	0.911098106261497\\
208000000000	0.910267861312828\\
209000000000	0.909436300555542\\
210000000000	0.90859920265211\\
211000000000	0.907754287817656\\
212000000000	0.906904951379362\\
213000000000	0.906050498406784\\
214000000000	0.905198437332838\\
215000000000	0.9043433986692\\
216000000000	0.903490323820893\\
217000000000	0.902639795371257\\
218000000000	0.901791999773673\\
219000000000	0.900946349560363\\
220000000000	0.900101234677333\\
221000000000	0.899259679730762\\
222000000000	0.898411068412391\\
223000000000	0.897555588357972\\
224000000000	0.896697130427537\\
225000000000	0.895829639832946\\
226000000000	0.894952855138363\\
227000000000	0.894065522237047\\
228000000000	0.893175375101264\\
229000000000	0.892278221954461\\
230000000000	0.891379447384143\\
231000000000	0.890487215259612\\
232000000000	0.88959005329505\\
233000000000	0.88869080769154\\
234000000000	0.88779274033184\\
235000000000	0.886894010646707\\
236000000000	0.885993695235679\\
237000000000	0.88509181210408\\
238000000000	0.88418904888251\\
239000000000	0.883286260355773\\
240000000000	0.882383886610962\\
241000000000	0.881479176858301\\
242000000000	0.880576379513805\\
243000000000	0.879668957051326\\
244000000000	0.878751900234524\\
245000000000	0.877826040658455\\
246000000000	0.87688885077695\\
247000000000	0.875947219758429\\
248000000000	0.874990936085091\\
249000000000	0.874034317661441\\
250000000000	0.873069155562889\\
251000000000	0.872108061487285\\
252000000000	0.871147710154584\\
253000000000	0.870188766942291\\
254000000000	0.869227639854351\\
255000000000	0.868272882478816\\
256000000000	0.86731758667761\\
257000000000	0.866360076870006\\
258000000000	0.865392083230753\\
259000000000	0.864425462590461\\
260000000000	0.863449889413958\\
261000000000	0.86247853878887\\
262000000000	0.861500827449939\\
263000000000	0.860520233646008\\
264000000000	0.859542399481005\\
265000000000	0.858561667219417\\
266000000000	0.857577444051514\\
267000000000	0.856588198089916\\
268000000000	0.855594001908659\\
269000000000	0.854602322069385\\
270000000000	0.853609654370666\\
271000000000	0.852622355379282\\
272000000000	0.851634279646826\\
273000000000	0.850642982165272\\
274000000000	0.849644766483252\\
275000000000	0.848633558517872\\
276000000000	0.84761014745693\\
277000000000	0.84658115360841\\
278000000000	0.84553934957244\\
279000000000	0.844494782627731\\
280000000000	0.843447699341039\\
281000000000	0.842401278640879\\
282000000000	0.841365286009085\\
283000000000	0.84033744238555\\
284000000000	0.839315072899685\\
285000000000	0.838299172345849\\
286000000000	0.837285507563165\\
287000000000	0.836270136860573\\
288000000000	0.835244667465849\\
289000000000	0.834212950891874\\
290000000000	0.83316938834364\\
291000000000	0.832120440824827\\
292000000000	0.831062328752937\\
293000000000	0.829996705199585\\
294000000000	0.828924780888792\\
295000000000	0.827847470900793\\
296000000000	0.826765688948175\\
297000000000	0.825686364519487\\
298000000000	0.824606479536789\\
299000000000	0.823526707874698\\
300000000000	0.82245573774235\\
301000000000	0.821468013319476\\
302000000000	0.820530170561685\\
303000000000	0.819654381634275\\
304000000000	0.818850218293886\\
305000000000	0.818123456514711\\
306000000000	0.817475634413336\\
307000000000	0.816904389144143\\
308000000000	0.816407861915681\\
309000000000	0.815970066479684\\
};
\end{axis}

\end{tikzpicture}%
\label{fig:ParametersbAgainstCarrierFrequency}}
			\caption{Optimal parameters of $\mathcal{E}_\N(a,b)$.}
			\label{fig:ParametersabAgainstCarrierFrequency}
		\end{figure}
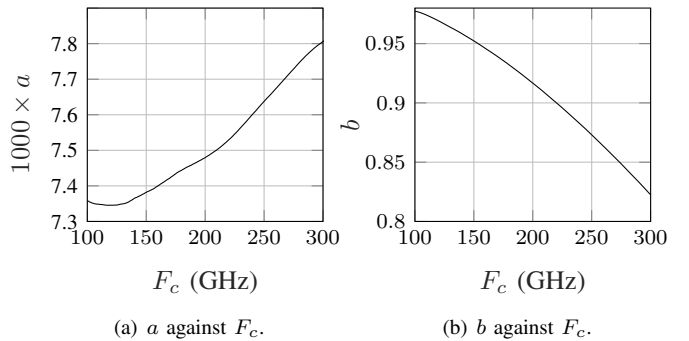

		\noindent
		By plotting the subsequent empirical autocorrelation function $\gamma(j)=\mathbb{E}[\alpha_n \alpha_{n-j}^*]$ of the generated $\underline{\alpha}$, see \figref{fig:MapTR3GPPToExp}, we observe a bimodal behavior. For low values of $j$, $\gamma(j)$ follows an exponential decrease like $e^{-a|j|}$ with $a\in\mathbb{R}_+$. For middle/large values of $j$, $\gamma(j)$ follows a linear decrease. The slope is low enough to consider a simple floor limit $\gamma(j)\approx b$ with $b\in\mathbb{R}$. From these observations, then, we propose the \textit{exponential model} $\mathcal{E}_\N(a,b)$:
		\begin{equation}
			\gamma_E(j) \eqdef \frac{e^{-a|j|} + \frac{b}{1-b}}{1+\frac{b}{1-b}}.
			\label{eq:ExponentialModel}
		\end{equation}
		The couple $(a,b)$ is found by minimizing the mean square error between $\underline{\gamma}$ and $\underline{\gamma}_E$. \figref{fig:MapTR3GPPToExp} exhibits $\underline{\gamma}$ and $\underline{\gamma}_E$ with the said parameters for several values of $F_c$. The evolution of $a,b$ as functions of $F_c$ are displayed in \figref{fig:ParametersaAgainstCarrierFrequency} and \figref{fig:ParametersbAgainstCarrierFrequency}. We observe that $a$ does not change a lot with $F_c$ compared to the change of $b$ with $F_c$. Therefore, an increase in the carrier frequency mainly involves a negative offset in the autocorrelation function which corresponds to a decrease in the correlation between the phase noise samples. This is consistent with the hypothesis of an uncorrelated phase noise model for very high frequencies proposed in \cite{Bicais2019}.

\section{Wiener filter\label{sec:WienerFilter}}
		This section presents the Wiener filter for tracking the phase noise when using the exponential model of the phase noise autocorrelation. The series of calculations is a tedious process that is not of a strong interest for the current paper. Therefore, here after, we only provide some clues and main results for mathematical derivations. The interested reader can refer to \cite{JCSarxiv} to get the whole demonstration. 
	\subsection{Basics\label{sec:WienerFilter_Basics}}
		From \cite{Oppenheim2010}, the Wiener filter provides an estimate of $\alpha_n$:
		\begin{equation}
			\hat{\alpha}_n = \underline{w}_n^T\tilde{\underline{\alpha}}_p,
			\label{eq:EstimateAlphaVector}
		\end{equation}
		where $\underline{w}_n \eqdef [w_{n,1} \dots w_{n,\NP}]$ are the filter coefficients and $\tilde{\underline{\alpha}}_p \eqdef [\tilde{\alpha}_{p_1} \dots \tilde{\alpha}_{p_{\NP}}]$ are the LS estimates on the pilot positions. The filter error is defined as $e_n \eqdef \alpha_n-\hat{\alpha}_n$ and the associated cost function is defined as:
		\begin{equation}
			J_n \eqdef \mathbb{E}[|e_n|^2].
			\label{eq:LocalCostFunction}
		\end{equation}
		The optimal filter coefficients $\underline{\hat{w}}_n$ are obtained by minimizing $J_n$ through a derivative calculation. It can be shown \cite{Oppenheim2010} that this amounts to:
		\begin{equation}
			\hat{\underline{w}}_n = R^{-1} \underline{\gamma}_n,
			\label{eq:WienerFilterCoefficients}
		\end{equation}
		where $R$ is the \textit{pilot autocorrelation matrix} of size $\NP\times\NP$ such that $R_{i,j} = \gamma(p_j-p_i)$ and $\underline{\gamma}_n$ is the \textit{autocorrelation vector} of size $\NP\times 1$ between a phase noise sample $\alpha_n$ (position $n$) and all the pilot positions $p_1,\dots,p_{\NP}$, such that the $j^{th}$ coordinate of $\underline{\gamma}_n$ is $\gamma(n-p_j)$. The autocorrelation function is then necessary to derive the analytical form of the Wiener filter coefficients.

	\subsection{Autocorrelation vector}
		We use the exponential model $\mathcal{E}_\N(a,b)$ to replace $\underline{\gamma}$ in \eqref{eq:WienerFilterCoefficients}. This makes the correlation vector become:
		\begin{equation}
			\underline{\gamma}_n =	\frac{1}{1+c}\left(\left[
									\begin{array}{c}
										e^{-a|n-p_1|}\\
										e^{-a|n-p_2|}\\
										\vdots \\
										e^{-a|n-p_{\NP}|}
									\end{array}
								\right]
								+
								c\left[
										\begin{array}{c}
											1\\
											1\\
											\vdots \\
											1
										\end{array}
									\right]
								\right).
			\label{eq:CorrelationVector}
		\end{equation}
		with $c\eqdef \frac{b}{1-b}$. Given \eqref{eq:pj}, it is possible to remove the dependence in the pilot position $p_j$ which will be convenient in the rest of the paper. Defining $\lambda\eqdef e^{-a\Delta}$, it comes that:
		\begin{equation}
			e^{-a|n-p_j|} = \left\{
						\begin{array}{cc}
							e^{-a(p_1-n)}\lambda^{j-1} & n \leq p_j \\
							e^{-a(n-p_1)}\lambda^{1-j} & \text{otherwise}
						\end{array}
					\right. .
		\end{equation}	
	\subsection{Pilot autocorrelation matrix}
		We can show that the autocorrelation matrix $R$ is the following sum:
		\begin{equation}
			R = \frac{c}{1+c}\Bigl(\frac{1}{c}A_{\lambda}+J_{\NP}\Bigr),
			\label{eq:CorrelationMatrixForExponential}
		\end{equation}
		with:
		\begin{equation}
			A_{\lambda} \eqdef \left[
						\begin{array}{cccc}
							1           	& \lambda         & \dots  & \lambda^{\NP-1} \\
							\lambda         & 1 		  & \dots  & \lambda^{\NP-2} \\
							\vdots 		& \vdots    	  & \ddots & \vdots \\
							\lambda^{\NP-1} & \lambda^{\NP-2} & \dots  & 1
						\end{array}
					\right], 
			\label{eq:Alambda}
		\end{equation}
		and:
		\begin{equation}
			J_{\NP} \eqdef \left[
						\begin{array}{cccc}
							1      & 1      & \dots  & 1 \\
							1      & 1      & \dots  & 1 \\
							\vdots & \vdots & \ddots & \vdots \\
							1      & 1      & \dots  & 1
						\end{array}
					\right].
		\end{equation}
		Defining the all-ones column vector $\underline{u}$ of length $\NP$, it comes that $J_{\NP} = \underline{u}\underline{u}^T$. From \eqref{eq:CorrelationMatrixForExponential}, we can say that:
		\begin{equation}
			R^{-1} = \frac{1+c}{c}\Bigl(\frac{1}{c}A_{\lambda}+J_{\NP}\Bigr)^{-1},
		\end{equation}
		Now, we make use of the Sherman-Morrison formula from \href{https://en.wikipedia.org/wiki/Sherman\%E2\%80\%93Morrison_formula}{\textit{weblink}} that leads to:
		\begin{equation}
			R^{-1} = (1+c)\Biggl(A_{\lambda}^{-1} - c\frac{A_{\lambda}^{-1} J_{\NP} A_{\lambda}^{-1}}{1+c\underline{u}^T A_{\lambda}^{-1} \underline{u}}\Biggr).
			\label{eq:SMF}
		\end{equation}
		According to \cite{StackExch} we can invert invert $A_{\lambda}$ s.t.:
		\begin{equation}
			A_{\lambda}^{-1} = \frac{X_{\lambda}}{1-\lambda^2},
			\label{eq:InvertingA}
		\end{equation}
		with:
		\begin{equation}
			X_{\lambda} \eqdef \left[
						\begin{array}{ccccccc}
							1	& -\lambda    	& 0       	& \dots  	& \dots   	& \dots   	& 0\\
							-\lambda& 1+\lambda^2 	& -\lambda    	& 0      	&         	&         	& \vdots\\
							0      	& -\lambda    	& 1+\lambda^2 	& -\lambda	& 0       	&         	& \vdots\\
							\vdots 	& \ddots	& \ddots  	& \ddots 	& \ddots	& \ddots	& \vdots \\
							\vdots 	&         	& 0       	& -\lambda	& 1+\lambda^2 	& -\lambda    	& 0 \\       
							\vdots 	&         	&         	& 0      	& -\lambda    	& 1+\lambda^2 	& -\lambda \\       
							0      	& \dots   	& \dots   	& \dots  	& 0       	& -\lambda    	& 1
						\end{array}
					\right].
		\end{equation}
		Now, we compute \eqref{eq:SMF} step-by-step. First of all, we can show that:
		\begin{equation}
			A_{\lambda}^{-1}J_{\NP} = \frac{1}{1+\lambda}\left[
									\begin{array}{ccc}
										1         & \dots & 1 \\
										1-\lambda & \dots & 1-\lambda \\
										\vdots    &       & \vdots \\
										1-\lambda & \dots & 1-\lambda \\
										1         & \dots & 1
									\end{array}
								\right].
		\end{equation}
		This leads to the calculation of the numerator of the fraction in \eqref{eq:SMF}:
		\begin{equation}
			A_{\lambda}^{-1}J_{\NP} A_{\lambda}^{-1} = \frac{Y_{\lambda}}{(1+\lambda)^2},
		\end{equation}
		with:
		\begin{equation}
			Y_{\lambda} \eqdef \left[
						\begin{array}{ccccc}
							1         & 1-\lambda     & \dots & 1-\lambda     & 1\\
							1-\lambda & (1-\lambda)^2 & \dots & (1-\lambda)^2 & 1-\lambda\\
							\vdots    & \vdots        &       & \vdots        & \vdots \\ 
							1-\lambda & (1-\lambda)^2 & \dots & (1-\lambda)^2 & 1-\lambda\\
							1         & 1-\lambda     & \dots & 1-\lambda     & 1
						\end{array}
					\right].
		\end{equation}
		After that, we focus on the denominator of the fraction in \eqref{eq:SMF}. First, we compute:
		\begin{equation}
			A_{\lambda}^{-1}\underline{u} = \frac{1}{1+\lambda}\left[
										\begin{array}{c}
											1\\
											1-\lambda\\
											\vdots\\
											1-\lambda\\
											1
										\end{array}
									\right].
		\end{equation}
		Then, it comes that:
		\begin{equation}
			\underline{u}^T A_{\lambda}^{-1}\underline{u} =  \frac{2\lambda+(1-\lambda)\NP}{1+\lambda}.
		\end{equation}
		Merging the result for the numerator and the denominator provides the following formula for the fraction:
		\begin{equation}
			c\frac{A_{\lambda}^{-1} J_{\NP} A_{\lambda}^{-1}}{1+c\underline{u}^T A_{\lambda}^{-1} \underline{u}} = \frac{cY_{\lambda}}{(1+\lambda)\bigl(1+\lambda + 2\lambda c + (1-\lambda)c\NP\bigr)}.
		\end{equation}
		Now, we deduce the equivalent form of $R^{-1}$ in \eqref{eq:SMF}:
		\begin{equation}
			R^{-1} = \frac{1+c}{1+\lambda}\Biggl( \frac{X_{\lambda}}{1-\lambda} - \frac{cY_{\lambda}}{1+\lambda + 2\lambda c + (1-\lambda)c\NP}\Biggr).
			\label{eq:InvertOfR}
		\end{equation}
		
	\subsection{Wiener filter coefficients}
		We inject \eqref{eq:CorrelationVector} and \eqref{eq:InvertOfR} in \eqref{eq:WienerFilterCoefficients} to obtain a closed form of the Wiener filter coefficients. To this end, we inject the expression of $R^{-1}$ from the previous section which results in $\hat{\underline{w}}_n = \hat{\underline{w}}_n^{(X)} - \hat{\underline{w}}_n^{(Y)}$ with:
		\begin{align}
			\hat{\underline{w}}_n^{(X)} \eqdef & \frac{1+c}{(1+\lambda)(1-\lambda)}X_{\lambda}\underline{\gamma}_n,\\
			\hat{\underline{w}}_n^{(Y)} \eqdef & \frac{c(1+c)}{(1+\lambda)\bigl(1+\lambda + 2\lambda c + (1-\lambda)c\NP\bigr)}Y_{\lambda}\underline{\gamma}_n.
		\end{align}
		Firstly, in the formula of $\hat{\underline{w}}_n^{(X)}$ we can show that:
		\begin{align}
			&(1+c)X_{\lambda}\underline{\gamma}_n = \Biggl(c(1-\lambda)\left[
										\begin{array}{c}
											1 \\
											1-\lambda\\
											\vdots\\
											1-\lambda\\
											1
										\end{array}
									\right]\\
								+&	\left[
										\begin{array}{c}
											e^{-a|n-p_1|}-\lambda e^{-a|n-p_2|} \\
											(1+\lambda^2) e^{-a|n-p_2|} -\lambda (e^{-a|n-p_1|} + e^{-a|n-p_3|}) \\
											\vdots\\
											(1+\lambda^2) e^{-a|n-p_j|} -\lambda (e^{-a|n-p_{j-1}|} + e^{-a|n-p_{j+1}|}) \\
											\vdots\\
											(1+\lambda^2) e^{-a|n-p_{\NP-1}|} -\lambda (e^{-a|n-p_{\NP-2}|} + e^{-a|n-p_{\NP}|}) \\
											e^{-a|n-p_{\NP}|}-\lambda e^{-a|n-p_{\NP-1}|}
										\end{array}
									\right]\Biggr),
		\end{align}
		where the row with $p_j$ is the $j^{\text{th}}$ row of the matrix. Therefore:
		\begin{align}
			&\hat{\underline{w}}_n^{(X)} = \frac{1}{1-\lambda^2}\nonumber\\
								&\times\left[
									\begin{array}{c}
										e^{-a|n-p_1|}-\lambda e^{-a|n-p_2|} \\
										(1+\lambda^2) e^{-a|n-p_2|} -\lambda (e^{-a|n-p_1|} + e^{-a|n-p_3|}) \\
										\vdots\\
										(1+\lambda^2) e^{-a|n-p_j|} -\lambda (e^{-a|n-p_{j-1}|} + e^{-a|n-p_{j+1}|}) \\
										\vdots\\
										(1+\lambda^2) e^{-a|n-p_{\NP-1}|} -\lambda (e^{-a|n-p_{\NP-2}|} + e^{-a|n-p_{\NP}|}) \\
										e^{-a|n-p_{\NP}|}-\lambda e^{-a|n-p_{\NP-1}|}
									\end{array}
								\right]\nonumber\\
							&+\frac{c}{1+\lambda}\left[
										\begin{array}{c}
											1 \\
											1-\lambda\\
											\vdots\\
											1-\lambda\\
											1
										\end{array}
									\right].
			\label{eq:wnX}
		\end{align}
		Secondly, in the formula of $\hat{\underline{w}}_n^{(Y)}$ it can be shown that:
		\begin{equation}
			(1+c)Y_{\lambda}\underline{\gamma}_n = \Bigl(\beta_n+\rho_{\lambda}\Bigr)\left[
													\begin{array}{c}
														1\\
														1-\lambda\\
														\vdots\\
														1-\lambda\\
														1
													\end{array}
												\right],
		\end{equation}
		with:
		\begin{align}
			\beta_n \eqdef & \sum_{j=1}^{\NP}e^{-a|n-p_j|} -\lambda \sum_{j=2}^{\NP-1}e^{-a|n-p_j|},\\
			\rho_{\lambda} \eqdef & \lambda c (2-\NP) + c\NP.
		\end{align}
		Therefore:
		\begin{equation}
			\hat{\underline{w}}_n^{(Y)} = \frac{c\bigl(\beta_n+\rho_{\lambda}\bigr)}{(1+\lambda)(1+\lambda+\rho_{\lambda})}\left[
																			\begin{array}{c}
																				1\\
																				1-\lambda\\
																				\vdots\\
																				1-\lambda\\
																				1
																			\end{array}
																		\right].
			\label{eq:wnY}
		\end{equation}
		Now, using \eqref{eq:wnX} and \eqref{eq:wnY}, we obtain $\hat{\underline{w}}_n$:
		\begin{align}
			&\hat{\underline{w}}_n = \frac{c\bigl(1+\lambda-\beta_n\bigr)}{(1+\lambda)(1+\lambda+\rho_{\lambda})}\left[
																				\begin{array}{c}
																					1\\
																					1-\lambda\\
																					\vdots\\
																					1-\lambda\\
																					1
																				\end{array}
																			\right]\nonumber\\
			&+\frac{1}{1-\lambda^2}\nonumber\\
								&\times\left[
									\begin{array}{c}
										e^{-a|n-p_1|}-\lambda e^{-a|n-p_2|} \\
										(1+\lambda^2) e^{-a|n-p_2|} -\lambda (e^{-a|n-p_1|} + e^{-a|n-p_3|}) \\
										\vdots\\
										(1+\lambda^2) e^{-a|n-p_j|} -\lambda (e^{-a|n-p_{j-1}|} + e^{-a|n-p_{j+1}|}) \\
										\vdots\\
										(1+\lambda^2) e^{-a|n-p_{\NP-1}|} -\lambda (e^{-a|n-p_{\NP-2}|} + e^{-a|n-p_{\NP}|}) \\
										e^{-a|n-p_{\NP}|}-\lambda e^{-a|n-p_{\NP-1}|}
									\end{array}
								\right]. \label{eq:WienerFilterCoefficientsRefinement}
		\end{align}
		
		From this analytical expression, we are given the possibility to study the behavior of the Wiener filter when changing the parameters, e.g., $a,b$ (or $F_c$) or the PT-RS spacing $\Delta$. As this is not the target of the work presented here and given the limited size of the paper, we focus on the PT-RS selection only and we let this analysis to the interested reader.
		
\section{Cost function derivation\label{sec:CostFunctionDerivation}}
	In this section, based on the formula of the Wiener filter coefficients \eqref{eq:WienerFilterCoefficientsRefinement}, we construct a global cost function $J$ stemming from the local cost functions $J_n$ defined in \eqref{eq:LocalCostFunction} that explicitly includes the PT-RS spacing $\Delta$ (or the equivalent quantity $\lambda$). Similarly to the previous section, the mathematical details are available in \cite{JCSarxiv}. Then, we show the behavior of the said function $J$ as a function of $a,b,F_c,\Delta$.
	
	\subsection{Cost function\label{sec:CostFunction}}
		We consider as the global cost function the sum of the local cost functions:
		\begin{equation}
			J = \sum_{n=1}^\N J_n.
			\label{eq:GlobalCostFunctionDefinition}
		\end{equation}
		By developing the local cost function $J_n$ from \eqref{eq:LocalCostFunction}, we obtain:
		\begin{equation}
			J_n = 1 - 2\underline{w}_n^T\mathcal{R}\{\mathbb{E}[\alpha_n^* \underline{\tilde{\alpha}}_p]\} + \underline{w}_n^T\mathbb{E}[\underline{\tilde{\alpha}}_p \underline{\tilde{\alpha}}_p^\dagger] \underline{w}_n.
		\end{equation}
		Then it comes that the global cost function is:
		\begin{equation}
			J = \N-\sum_{n=1}^\N \underline{w}_n^T \underline{\gamma}_n.
			\label{eq:GlobalCostFunction}
		\end{equation}
		From \eqref{eq:LocalCostFunction} and \eqref{eq:GlobalCostFunctionDefinition}, we observe that $J\geq 0$. In addition, by replacing $\underline{w}_n$ with \eqref{eq:WienerFilterCoefficients}, we see that $\underline{w}_n^T \underline{\gamma}_n$ is a quadratic form, i.e, it is positive. Therefore, $J$ is upper and lower bounded such that $0\leq J\leq \N$.
		
	\subsection{Derivation}
		We need to expand $J$ as $J(\lambda)$, an explicit function of $\lambda$. To this end we need to compute the scalar product $\underline{w}_n^T \underline{\gamma}_n$ for all $n$. We expand $\underline{\gamma}_n$ as:
		\begin{equation}
			\underline{\gamma}_n = \frac{1}{1+c}\underline{\gamma}_n^{(0)} + \frac{c}{1+c}\underline{u},
		\end{equation}
		where $\underline{u}$ is the full-one column vector of length $\NP$ (see previous Section) and $\gamma_{n,j}^{(0)} \eqdef e^{-a|n-p_j|}$. 
		
		For this part, we focus on computing $\underline{w}_n^T \underline{\gamma}_n^{(0)}$. As $\underline{w}_n$ is a sum of two terms, we compute the product between the transpose of each of these terms and $\underline{\gamma}_n^{(0)}$. First of all, we can show that:
		\begin{equation}
			\left[
				\begin{array}{c}
					1\\
					1-\lambda\\
					\vdots\\
					1-\lambda\\
					1
				\end{array}
			\right]^T \cdot\underline{\gamma}_n^{(0)} = \beta_n.
		\end{equation}
		Secondly, we focus on the second term defining $\underline{v}_n$ as:
		\begin{eqnarray}
			v_{n,j} & = & (1+\lambda^2) e^{-a|n-p_j|} -\lambda (e^{-a|n-p_{j-1}|} + e^{-a|n-p_{j+1}|}), \nonumber \\
			v_{n,1} & = & e^{-a|n-p_1|}-\lambda e^{-a|n-p_2|},\\
			v_{n,\N} & = & e^{-a|n-p_{\NP}|}-\lambda e^{-a|n-p_{\NP-1}|}. \nonumber
		\end{eqnarray}
		It can be shown that:
		\begin{align}
			\underline{v}_n^T\underline{\gamma}_n^{(0)} = & \sum_{j=1}^{\NP} \biggl(e^{-a|n-p_j|}\biggr)^2 +\sum_{j=2}^{\NP-1} \biggl(\lambda e^{-a|n-p_j|}\biggr)^2 \nonumber\\
										 & -2\lambda\sum_{j=1}^{\NP-1} e^{-a|n-p_j|}e^{-a|n-p_{j+1}|},
		\end{align}
		which amounts to:
		\begin{align}
			\underline{v}_n^T\underline{\gamma}_n^{(0)} = & \sum_{j=1}^{\NP-1} \biggl(e^{-a|n-p_j|}-\lambda e^{-a|n-p_{j+1}|}\biggr)^2 \nonumber \\
										 & +(1-\lambda^2)e^{-2a|n-p_{\NP}|}.
		\end{align}			
		Let's consider three cases to compute the sum: $n\leq p_1$, $p_1<n<p_{\NP}$, $p_{\NP}\leq n$. For a given $n$, we introduce $\KP(n)$ the number of pilots before the index $n$:
		\begin{equation}
			\KP(n) \eqdef |\{j | p_j \leq n\}|.
		\end{equation}
		
		\noindent
		\underline{Case 1:} $n\leq p_1$\\
		Here:
		\begin{align}
			\underline{v}_n^T\underline{\gamma}_n^{(0)} = (1-\lambda^2)e^{-2a(p_1-n)} \Bigl( 1-(1-\lambda)\lambda^{2(\NP-1)} \Bigr)
		\end{align}
		
		\noindent
		\underline{Case 2:} $p_1<n<p_{\NP}$\\
		Here:
		\begin{align}
			\underline{v}_n^T\underline{\gamma}_n^{(0)} =& (1-\lambda^2) e^{-2a(p_1-n)} \lambda^{2\KP(n)} \\
										& + \lambda^2 \biggl(e^{-a(n-p_1)}\lambda^{-\KP(n)} - e^{-a(p_1-n)}\lambda^{\KP(n)}\biggr)^2 \nonumber.
		\end{align}
		
		\noindent
		\underline{Case 3:} $p_{\NP}\leq n$\\
		Here:
		\begin{align}
			\underline{v}_n^T\underline{\gamma}_n^{(0)} = (1-\lambda^2)e^{-2a(n-p_1)}\lambda^{2-2\NP}.
		\end{align}
		Now, we focus on computing $\underline{w}_n^T \underline{u}$. Actually, this product amounts to sum all the coordinates of $\underline{w}_n(\lambda)$. We can demonstrate that:
		\begin{equation}
			\underline{w}_n^T \underline{u} = \frac{1+\lambda-\beta_n}{(1+\lambda)(1+\lambda+\rho_{\lambda})}\rho_{\lambda} + \frac{1}{1-\lambda^2} \sum_{j=1}^{\NP} v_{n,j}.
		\end{equation}
		It appears that:
		\begin{align}
			\frac{1}{1-\lambda^2} \sum_{j=1}^{\NP} v_{n,j} = \frac{1}{1+\lambda}\Bigl(& \lambda\bigl(e^{-a|n-p_1|}+e^{-a|n-p_{\NP}|}\bigr) \nonumber \\
													   & + (1-\lambda)\sum_{j=1}^{\NP}e^{-a|n-p_j|}\Bigr),
		\end{align}
		where we recognize the component $\beta_n$ such that:
		\begin{equation}
			\frac{1}{1-\lambda^2} \sum_{j=1}^{\NP} v_{n,j} = \frac{1}{1+\lambda}\beta_n,
		\end{equation}
		therefore:
		\begin{equation}
			\underline{w}_n^T \underline{u} = \frac{1+\lambda-\beta_n}{(1+\lambda)(1+\lambda+\rho_{\lambda})}\rho_{\lambda} + \frac{1}{1+\lambda}\beta_n.
		\end{equation}
		These previous calculations allows us to obtain:
		\begin{align}
			\underline{w}_n^T \underline{\gamma}_n = 	& c\frac{
											\beta_n \bigl(
														2(1+\lambda)-\beta_n
													   \bigr)
											+ \rho_{\lambda}(1+\lambda)}
											{
											(1+c)(1+\lambda)(1+\lambda+\rho_{\lambda})
											} \nonumber \\
										& + \frac{1}{(1+c)(1-\lambda^2)} \underline{v}_n^T \underline{\gamma}_n^{(0)}.
		\end{align}

		This provides the expression of the cost function:
		\begin{align}
			J(\lambda) = 	& \N - \frac{c}{(1+c)(1+\lambda)(1+\lambda+\rho)} \nonumber\\
				& \times \Biggl( 2(1+\lambda) \sum_{n=1}^\N \beta_n - \sum_{n=1}^\N \beta_n^2 + \rho(1+\lambda)\N \Biggr) \nonumber\\
				& -\frac{1}{(1+c)(1-\lambda^2)} \sum_{n=1}^\N\underline{u}_n^T \underline{\gamma}_n^{(0)}.
		\end{align}		
		To derive this calculation, we need to expand $\beta_n$ and the sums of $\beta_n$ and $\beta_n^2$.
	\subsection{Derivation of $\beta_n$}
		We have to distinguish according to the value of $n$  in comparison with the pilots position.	
			
		\noindent
		\underline{Case 1:} $n< p_1$\\		
		Here, we can show that:
		\begin{align}
			\beta_n = e^{-a(p_1-n)}(1+\lambda).
		\end{align}		
		
		\noindent
		\underline{Case 2:} $p_1\leq n<p_{\NP}$\\
		Here, we can show that:
		\begin{align}
			\beta_n = e^{-a(n-p_1)}\lambda^{1-\KP(n)} + e^{-a(p_1-n)}\lambda^{\KP(n)}.
		\end{align}		
		
		\noindent
		\underline{Case 3:} $p_{\NP}\leq n$\\
		Here, we can show that:
		\begin{align}
			\beta_n = e^{-a(n-p_1)}\lambda^{1-\NP}(1+\lambda).
		\end{align}
	
	\subsection{Derivation of $\sum_n\beta_n$}
		The sum over $n$ is split into three sums as follows:
		\begin{align}
			\sum_{n=1}^\N \beta_n = 	& \sum_{n=1}^{p_1-1} \beta_n \nonumber\\
								& + \sum_{n=p_1}^{p_{\NP}-1} \beta_n \nonumber\\
								& + \sum_{n=p_{\NP}}^{\N} \beta_n.
		\end{align}
		It can be shown that:
		\begin{align}
			\sum_{n=1}^{p_1-1} \beta_n = & (1+\lambda) \frac{e^{-a(p_1 -1)}-1}{1-e^a}, \\
			\sum_{n=p_{\NP}}^{\N} \beta_n = & (1+\lambda) \frac{1-\lambda^{1-\NP}e^{-a(\N+1-p_1)}}{1-e^{-a}}.
		\end{align}
		To compute the sum for $p_1 < n < p_{\NP}$, we consider a first split according to the expression of $\beta_n$:
		\begin{align}
			\sum_{n=p_1}^{p_{\NP}-1} \beta_n = 	& \lambda e^{a p_1} \sum_{n=p_1}^{p_{\NP}-1} e^{-an} \lambda^{-\KP(n)} \nonumber\\
									& + e^{-a p_1} \sum_{n=p_1}^{p_{\NP}-1} e^{an}\lambda^{\KP(n)},
		\end{align}
		then, we split each of both sums into sub-sums where each of them goes from one pilot position $p_j$ to the next one $p_{j+1}-1$ such that:
		\begin{align}
			\sum_{n=p_1}^{p_{\NP}-1} e^{-an} \lambda^{-\KP(n)} = & \sum_{j=1}^{\NP-1}\sum_{n=p_j}^{p_{j+1}-1} e^{-an} \lambda^{-\KP(n)},\\
			\sum_{n=p_1}^{p_{\NP}-1} e^{an} \lambda^{\KP(n)} = & \sum_{j=1}^{\NP-1}\sum_{n=p_j}^{p_{j+1}-1} e^{an} \lambda^{\KP(n)}.
		\end{align}
		The rationale behind is that for any $p_j\leq n < p_{j+1}$ we observe that $\KP(n)=j$. Therefore:
		\begin{align}
			\sum_{n=p_1}^{p_{\NP}-1} e^{-an} \lambda^{-\KP(n)} = & \sum_{j=1}^{\NP-1} \lambda^{-j} \sum_{n=p_j}^{p_{j+1}-1} e^{-an},\\
			\sum_{n=p_1}^{p_{\NP}-1} e^{an} \lambda^{\KP(n)} = & \sum_{j=1}^{\NP-1} \lambda^j \sum_{n=p_j}^{p_{j+1}-1} e^{an}.
		\end{align}
		Now it comes that:
		\begin{align}
			\sum_{n=p_1}^{p_{\NP}-1} e^{-an} \lambda^{-\KP(n)} = & e^{-a p_1} (\NP-1)\frac{1-\lambda}{\lambda(1-e^{-a})},\\
			\sum_{n=p_1}^{p_{\NP}-1} e^{an} \lambda^{\KP(n)} = & e^{a p_1} (\NP-1)\frac{\lambda-1}{1-e^a}.
		\end{align}
		This provides then the middle sum of $\beta_n$:
		\begin{align}
			\sum_{n=p_1}^{p_{\NP}-1} \beta_n = (\NP-1)(1-\lambda)\biggl(\frac{1}{1-e^{-a}} - \frac{1}{1-e^a}\biggr).
		\end{align}
		Therefore, the total sum is:
		\begin{align}
			\sum_{n=1}^\N \beta_n = 	& \frac{1+\lambda}{1-e^a} \Bigl( e^{-a(p_1 -1)}-1 \Bigr) \nonumber\\
								& + (\NP-1)(1-\lambda)\biggl(\frac{1}{1-e^{-a}} - \frac{1}{1-e^a}\biggr) \nonumber\\ 
								& + \frac{1+\lambda}{1-e^{-a}} \Bigl(1-\lambda^{1-\NP}e^{-a(\N+1-p_1)} \Bigr).
		\end{align}
		This can also be written as:
		\begin{align}
			\sum_{n=1}^\N \beta_n = 	& \Bigl( \lambda(\NP-2)-\NP \Bigr) \frac{1+e^a}{1-e^a} \nonumber\\
								& \frac{1+\lambda}{1-e^a} \Bigl( e^{-a(p_1-1)} + e^{-a(\N-p_1)}\lambda^{1-\NP} \Bigr)
		\end{align}
	\subsection{Derivation of $\sum_n\beta_n^2$}
		Similarly, the sum over $n$ is split into three sums as follows:
		\begin{align}
			\sum_{n=1}^\N \beta_n^2 = 	& \sum_{n=1}^{p_1-1} \beta_n^2 \nonumber\\
								& + \sum_{n=p_1}^{p_{\NP}-1} \beta_n^2 \nonumber\\
								& + \sum_{n=p_{\NP}}^{\N} \beta_n^2.
		\end{align}
		It can be shown that:
		\begin{align}
			\sum_{n=1}^{p_1-1} \beta_n^2 = & (1+\lambda)^2 \frac{e^{-2a(p_1 -1)}-1}{1-e^{2a}}, \\
			\sum_{n=p_{\NP}}^{\N} \beta_n^2 = & (1+\lambda)^2 \frac{1-\lambda^{2(1-\NP)}e^{-2a(\N+1-p_1)}}{1-e^{-2a}}.
		\end{align}
		In the same manner as previously, we can show that:
		\begin{align}
			\sum_{n=p_1}^{p_{\NP}-1} \beta_n^2 = 2\lambda\Delta(\NP-1) + (1-\lambda^2)(\NP-1)\frac{e^{2a}+1}{e^{2a}-1}.
		\end{align}
		We can then merge these three sums together to obtain:
		\begin{align}
			\sum_{n=1}^\N \beta_n^2 = 	& (1+\lambda)\Bigl(\lambda(\NP-2)-\NP\Bigr)\frac{1+e^{2a}}{1-e^{2a}} \nonumber\\
								& + \frac{(1+\lambda)^2}{1-e^{2a}}\biggl( e^{-2a(p_1-1)} + \lambda^{2(1-\NP)}e^{-2a(\N-p_1)} \biggr) \nonumber\\
								& + 2\lambda\Delta(\NP-1)
		\end{align}
											
	\subsection{Derivation of $J_{\beta}(\lambda)$}
		We define the quantity:
		\begin{align}
			J_{\beta}(\lambda) \eqdef 2(1+\lambda) \sum_{n=1}^\N \beta_n - \sum_{n=1}^\N \beta_n^2,
		\end{align}
		which is part of the total cost function $J(\lambda)$. We can show that:
		\begin{align}
			J_{\beta}(\lambda) = 	& (1+\lambda)\Bigl(\lambda(\NP-2)-\NP\Bigr) \frac{(1-e^a)^2 + 6e^a}{1-e^{2a}} \nonumber\\
						& + \frac{(1+\lambda)^2}{1-e^a}	\biggl(
											2e^{-a(p_1-1)} + 2e^{-a(\N-p_1)}\lambda^{1-\NP} \nonumber\\
											& -\frac{e^{-2a(p_1-1)}}{1+e^a} - \frac{e^{-2a(\N-p_1)}\lambda^{2(1-\NP)}}{1+e^a}
										\biggr) \nonumber\\
						& -2\lambda\Delta(\NP-1)
		\end{align}
	
	\subsection{Derivation of $\sum_n \underline{v}_n^T \underline{\gamma}_n^{(0)}$}
		Similarly to the sum of $\beta_n$ we split the sum into three sums such that:
		\begin{align}
			\sum_{n=1}^\N \underline{v}_n^T \underline{\gamma}_n^{(0)} = & \sum_{n=1}^{p_1-1} \underline{v}_n^T \underline{\gamma}_n^{(0)} \nonumber\\
												& + \sum_{n=p_1}^{p_{\NP}-1} \underline{v}_n^T \underline{\gamma}_n^{(0)} \nonumber\\
												& + \sum_{n=p_{\NP}}^\N \underline{v}_n^T \underline{\gamma}_n^{(0)}.
		\end{align}
		We can show that:
		\begin{align}
			\sum_{n=1}^{p_1-1} \underline{v}_n^T \underline{\gamma}_n^{(0)} = 	& (1-\lambda^2)\Bigl( 1-(1-\lambda)\lambda^{2(\NP-1)} \Bigr) \nonumber \\
													& \times \frac{e^{-2a(p_1-1)}-1}{1-e^{2a}},
		\end{align}
		and that:
		\begin{align}
			\sum_{n=p_{\NP}}^\N  \underline{v}_n^T \underline{\gamma}_n^{(0)} = 	(1-\lambda^2)\frac{1-\lambda^{2(1-\NP)}e^{-2a(\N+1-p_1)}}{1-e^{-2a}}.
		\end{align}
		For the sum on middle values of $n$, we use the same method as for the sum on middle values of $\beta_n$, i.e.:
		\begin{align}
			\sum_{n=p_1}^{p_{\NP}-1} \underline{v}_n^T \underline{\gamma}_n^{(0)} = \sum_{j=1}^{\NP-1} \sum_{n=p_j}^{p_{j+1}-1} \underline{v}_n^T \underline{\gamma}_n^{(0)},
		\end{align}
		which gives:
		\begin{align}
			\sum_{n=p_1}^{p_{\NP}-1} \underline{v}_n^T \underline{\gamma}_n^{(0)} = 	& (1-\lambda^2)e^{-2a p_1} \sum_{j=1}^{\NP-1} \sum_{n=p_j}^{p_{j+1}-1} \lambda^{2\KP(n)}e^{2an} \nonumber\\
														& + \lambda^2 \sum_{j=1}^{\NP-1} \sum_{n=p_j}^{p_{j+1}-1} \Bigl( e^{-a(n-p_1)}\lambda^{-\KP(n)} \nonumber\\
														& - e^{-a(p_1-n)}\lambda^{\KP(n)} \Bigr)^2.
		\end{align}
		We can then show that:
		\begin{align}
			\sum_{n=p_1}^{p_{\NP}-1} \underline{v}_n^T \underline{\gamma}_n^{(0)} = (1-\lambda^2)(\NP-1)\frac{e^{2a}+1}{e^{2a}-1} - 2\lambda^2\Delta(\NP-1)
		\end{align}
		Finally, the total sum can be demonstrated to equal:
		\begin{align}
			\sum_{n=1}^\N \underline{v}_n^T \underline{\gamma}_n^{(0)} = & \frac{1-\lambda^2}{1-e^{2a}}	\Biggl(e^{-2a(p_1-1)} -\NP(e^{2a}+1) \nonumber\\
																	&  + \lambda^{2(1-\NP)}e^{-2a(\N-p_1)} \nonumber\\
																	& - (1-\lambda)\lambda^{2(\NP-1)}\Bigl( e^{-2a(p_1-1)}-1 \Bigr)
																\Biggr)\nonumber\\
												& -2\lambda^2\Delta(\NP-1)
		\end{align}
			
		At this stage, from the previous calculations, we are provided a one-dimensional closed form of the cost function $J(\lambda)$:	
		\begin{strip}
			\fbox{
				\begin{minipage}{\textwidth}
					\begin{align}
						J(\lambda) =  N - \frac{1}{1+c}	\textcolor{green}{\Biggl[} 	& \frac{c}{1+\lambda+\rho_{\lambda}}	\textcolor{blue}{\Biggl(} \rho_{\lambda}	\biggl(	\N - \frac{(1-e^a)^2+6e^a}{c(1-e^{2a})}	\biggr)\nonumber\\
														& \quad\quad\quad\quad\quad\quad\quad          			 + \frac{1+\lambda}{1-e^a} \textcolor{red}{\biggl(} 2e^{-a(p_1-1)} + 2e^{-a(\N-p_1)}\lambda^{1-\NP} \nonumber\\
														& \quad\quad\quad\quad\quad\quad\quad\quad\quad\quad\quad\quad 				 	   - \frac{e^{-2a(p_1-1)} + e^{-2a(\N-p_1)}\lambda^{2(1-\NP)}}{1+e^a} \textcolor{red}{\biggr)} \textcolor{blue}{\Biggr)}\nonumber\\
														& + \frac{1}{1-e^{2a}}	\textcolor{blue}{\biggl(} e^{-2a(p_1-1)} - \NP(e^{2a}+1) + \lambda^{2(1-\NP)}e^{-2a(\N-p_1)} \nonumber\\
														& \quad\quad\quad\quad\quad - (1-\lambda)\lambda^{2(\NP-1)}\Bigl(e^{-2a(p_1-1)}-1\Bigr) \textcolor{blue}{\biggr)} \nonumber\\
														& - \frac{\lambda}{1+\lambda}2\Delta(\NP-1)	\biggl( \frac{c}{1+\lambda+\rho_{\lambda}} + \frac{\lambda}{1-\lambda} \biggr) \textcolor{green}{\Biggr]} 
					\end{align}
				\end{minipage}
				}
		\end{strip}
		By identification, it can be shown that $J(\lambda)$ has a quasi-polynomial form such that:
		\begin{equation}
			J(\lambda) =  N - \frac{1}{1+c}	\frac{J^{(N)}(\lambda)}{J^{(D)}(\lambda)}.
			\label{eq:GlobalCostFunctionClosedForm}
		\end{equation}
		$J^{(N)}(\lambda)$ is defined with the monomials $\{j_k^{(N)}\}_k$ from Table \ref{tab:NumeratorMonomials} such that:
		\begin{align}
			J^{(N)}(\lambda) = \;& 	j_{2\NP+2}^{(N)} \lambda^{2\NP+2} + j_{2\NP+1}^{(N)} \lambda^{2\NP+1} \\
						& + j_{2\NP}^{(N)} \lambda^{2\NP} + j_{2\NP-1}^{(N)} \lambda^{2\NP-1} \nonumber\\
						& + j_{2\NP-2}^{(N)} \lambda^{2\NP-2} \nonumber\\
						& + j_3^{(N)} \lambda^3 + j_2^{(N)} \lambda^2 + j_1^{(N)} \lambda + j_0^{(N)} \nonumber\\
						& + j_{4-\NP}^{(N)} \lambda^{4-\NP} + j_{3-\NP}^{(N)} \lambda^{3-\NP} \nonumber\\
						& + j_{2-\NP}^{(N)} \lambda^{2-\NP} + j_{1-\NP}^{(N)} \lambda^{1-\NP} \nonumber\\
						& + j_{5-2\NP}^{(N)} \lambda^{5-2\NP} + j_{4-2\NP}^{(N)} \lambda^{4-2\NP} \nonumber\\
						& + j_{3-2\NP}^{(N)} \lambda^{3-2\NP} + j_{2-2\NP}^{(N)} \lambda^{2-2\NP} \nonumber\\
						& + \frac{\log \lambda}{a}\biggl( jj_3^{(N)} \lambda^3 + jj_2^{(N)} \lambda^2 + jj_1^{(N)} \lambda \biggr) \nonumber,
		\end{align}
		and:
		\begin{align}
			jj_3^{(N)} = & 2(1-\NP) \bigl( 1+c(2-\NP) \bigr),\\
			jj_2^{(N)} = & 2(1-\NP) \bigl( 1+c(\NP-1) \bigr), \nonumber\\
			jj_1^{(N)} = & 2(1-\NP)c. \nonumber
		\end{align}
		$J^{(D)}(\lambda)$ is defined with the monomials $\{j_k^{(D)}\}_k$ such that:
		\begin{equation}
			J^{(D)}(\lambda) = j_3^{(D)} \lambda^3 + j_2^{(D)} \lambda^2 + j_1^{(D)} \lambda + j_0^{(D)},
		\end{equation}
		and:
		\begin{align}
			j_3^{(D)} = & c(\NP-2)-1,\\
			j_2^{(D)} = & -(1+c\NP), \nonumber\\
			j_1^{(D)} = & -j_3^{(D)}, \nonumber\\
			j_0^{(D)} = & -j_2^{(D)}. \nonumber
		\end{align}
				
		\begin{table}[!h]
			\centering
			$$
			\begin{array}{|c|c|}
				\hline
				k & j_k^{(N)} \\
				\hline \hline
				2\NP+2 	& \frac{e^{-2a(p_1-1)}-1}{e^{2a}-1}\bigl( 1+c(2-\NP) \bigr) \\ \hline
				2\NP+1 	& 2c\frac{e^{-2a(p_1-1)}-1}{1-e^{2a}}(1-\NP) \\ \hline
				2\NP	& 2(1+c)\frac{e^{-2a(p_1-1)}-1}{1-e^{2a}} \\ \hline
				2\NP-1	& -j_{2\NP+1}^{(N)} \\ \hline
				2\NP-2	& \frac{e^{-2a(p_1-1)}-1}{e^{2a}-1}(1+c\NP) \\ \hline
				3	& \begin{array}{c}
						\Biggl( \bigl( 1+c(2-\NP) \bigr) \Bigl( \NP(e^{2a}+1) - e^{-2a(p_1-1)} \Bigr)  \\
							+ c(\NP-2)\Bigl( c\N(1-e^{2a}) - (1-e^a)^2 -6e^a \Bigr) \\
							- 2ce^{-a(p_1-1)}(e^a+1) + ce^{-2a(p_1-1)}
						\Biggr)/(1-e^{2a})
					\end{array} \\ \hline
				2	& \begin{array}{c}
						c^2\NP \Bigl( \frac{6e^a + (1-e^a)^2}{c(1-e^{2a})} - \N \Bigr) \\
						+ \frac{c}{1-e^a} \Bigl( \frac{e^{-2a(p_1-1)}}{e^a +1} - 2e^{-a(p_1-1)} \Bigr) \\
						+ \frac{1+c\NP}{1-e^{2a}}\Bigl( \NP(e^{2a}+1) - e^{-2a(p_1-1)} \Bigr)
					\end{array} \\ \hline
				1	& -j_3^{(N)} \\ \hline
				0	& -j_2^{(N)} \\ \hline
				4-\NP	& 2c\frac{e^{-a(\N-p_1)}}{e^a-1} \\ \hline
				3-\NP	& j_{4-\NP}^{(N)} \\ \hline
				2-\NP	& -j_{4-\NP}^{(N)} \\ \hline
				1-\NP	& -j_{4-\NP}^{(N)} \\ \hline
				5-2\NP	& \frac{e^{-2a(N-p_1)}}{e^{2a}-1}\bigl( 1+c(1-\NP) \bigr) \\ \hline
				4-2\NP	& \frac{e^{-2a(N-p_1)}}{e^{2a}-1}\bigl( 1+c(\NP-1) \bigr) \\ \hline
				3-2\NP	& -j_{5-2\NP}^{(N)} \\ \hline
				2-2\NP	& -j_{4-2\NP}^{(N)} \\
				\hline
			\end{array}
			$$
			\caption{Monomials of the numerator.}
			\label{tab:NumeratorMonomials}
		\end{table}
				
	
	\subsection{Influence of $a,b$ and $F_c$}
\begin{figure}[!h]
			\centering
			\subfigure[$J$ against $a$]{
%
%
\definecolor{mycolor1}{rgb}{0.00000,0.50000,1.00000}%
\definecolor{mycolor2}{rgb}{0.00000,1.00000,1.00000}%
\begin{tikzpicture}

\begin{axis}[%
	width=0.35\linewidth,
	height=0.12\textheight,
	font=\footnotesize,
	scale only axis,
	scaled ticks=false, 
	tick label style={/pgf/number format/fixed},
	xmin=0.007,
	xmax=0.008,
	xlabel style={font=\color{white!15!black}},
	xlabel={$1000\times a$},
	xtick={0.007,0.0072,0.0074,0.0076,0.0078,0.008},
	xticklabels={$7$,$7.2$,$7.4$,$7.6$,$7.8$,$8$},
	ymin=0,
	ymax=3,
	ylabel={$J$ (in \% of $N$)},
	ylabel style={font=\color{white!15!black},anchor=south west, at={(0.3,0.15)}},
	axis background/.style={fill=white},
	xmajorgrids,
	ymajorgrids,
	legend columns=2,
	legend style={legend cell align=left, align=left, draw=white!15!black, font=\tiny, anchor=south east, at={(1,1.1)}}
]
\addplot [color=mycolor1,mark=+,mark repeat=2]
  table[row sep=crcr]{%
0.007	2.32052353820114\\
0.0071	2.35309297778692\\
0.0072	2.38563887035604\\
0.0073	2.41816093162389\\
0.0074	2.450658878312\\
0.0075	2.48313242815651\\
0.0076	2.51558129992278\\
0.0077	2.54800521341553\\
0.0078	2.58040388949011\\
0.0079	2.6127770500637\\
0.008	2.64512441812558\\
};
\addlegendentry{$b = 0.8$}

\addplot [color=mycolor2,mark=o,mark repeat=2]
  table[row sep=crcr]{%
0.007	1.74039363941796\\
0.0071	1.76482073725057\\
0.0072	1.7892301750687\\
0.0073	1.81362173965954\\
0.0074	1.83799521856439\\
0.0075	1.86235040008687\\
0.0076	1.88668707330119\\
0.0077	1.91100502806104\\
0.0078	1.93530405500915\\
0.0079	1.95958394558322\\
0.008	1.98384449202597\\
};
\addlegendentry{$b = 0.85$}

\addplot [color=white!50!green,mark=x,mark repeat=2]
  table[row sep=crcr]{%
0.007	1.16026301579494\\
0.0071	1.1765477584632\\
0.0072	1.19282072793665\\
0.0073	1.20908178207308\\
0.0074	1.22533077923422\\
0.0075	1.24156757828838\\
0.0076	1.25779203861986\\
0.0077	1.27400402013116\\
0.0078	1.29020338325053\\
0.0079	1.30638998893701\\
0.008	1.32256369868616\\
};
\addlegendentry{$b = 0.9$}

\addplot [color=orange,mark=square,mark repeat=2]
  table[row sep=crcr]{%
0.007	0.580131773793413\\
0.0071	0.58827414996182\\
0.0072	0.596410639598788\\
0.0073	0.604541171633666\\
0.0074	0.612665675247659\\
0.0075	0.620784079874959\\
0.0076	0.628896315207517\\
0.0077	0.637002311197521\\
0.0078	0.645101998058206\\
0.0079	0.653195306269994\\
0.008	0.661282166580279\\
};
\addlegendentry{$b = 0.95$}

\end{axis}

\end{tikzpicture}%
}
			\subfigure[$J$ against $b$]{
%
%
\definecolor{mycolor1}{rgb}{0.00000,0.50000,1.00000}%
\definecolor{mycolor2}{rgb}{0.00000,1.00000,1.00000}%
\begin{tikzpicture}

\begin{axis}[%
	width=0.35\linewidth,
	height=0.12\textheight,
	font=\footnotesize,
	scale only axis,
	xmin=0.8,
	xmax=1,
	xlabel style={font=\color{white!15!black}},
	xlabel={$b$},
	ymin=0,
	ymax=3,
	ylabel={$J$ (in \% of $N$)},
	ylabel style={font=\color{white!15!black},anchor=south west, at={(0.3,0.15)}},
	axis background/.style={fill=white},
	xmajorgrids,
	ymajorgrids,
	legend columns=2,
	legend style={legend cell align=left, align=left, draw=white!15!black, font=\tiny, anchor=south east, at={(1,1.1)}}
]

\addplot [color=mycolor1,mark=+,mark repeat=4]
  table[row sep=crcr]{%
0.8	2.38563887035604\\
0.81	2.26635719760238\\
0.82	2.14707549057852\\
0.83	2.02779375042751\\
0.84	1.90851197824232\\
0.85	1.7892301750687\\
0.86	1.66994834190639\\
0.87	1.55066647971398\\
0.88	1.43138458940777\\
0.89	1.31210267186791\\
0.9	1.19282072793669\\
0.91	1.0735387584226\\
0.92	0.954256764101846\\
0.93	0.834974745719008\\
0.94	0.715692703988735\\
0.95	0.596410639598788\\
0.96	0.477128553209194\\
0.97	0.357846445455223\\
0.98	0.238564316947476\\
0.99	0.119282168273782\\
};
\addlegendentry{$a = 0.0072$}

\addplot [color=mycolor2,mark=o,mark repeat=4]
  table[row sep=crcr]{%
0.8	2.450658878312\\
0.81	2.32812621512576\\
0.82	2.20559351639598\\
0.83	2.08306078330998\\
0.84	1.96052801700415\\
0.85	1.83799521856439\\
0.86	1.71546238903052\\
0.87	1.59292952939729\\
0.88	1.4703966406176\\
0.89	1.34786372360471\\
0.9	1.2253307792342\\
0.91	1.10279780834522\\
0.92	0.980264811744458\\
0.93	0.857731790204097\\
0.94	0.735198744467469\\
0.95	0.612665675247659\\
0.96	0.490132583230374\\
0.97	0.367599469074809\\
0.98	0.245066333414912\\
0.99	0.122533176861117\\
};
\addlegendentry{$a = 0.0074$}

\addplot [color=white!50!green,mark=x,mark repeat=4]
  table[row sep=crcr]{%
0.8	2.51558129992278\\
0.81	2.38980252588666\\
0.82	2.26402371499933\\
0.83	2.13824486849454\\
0.84	2.0124659875524\\
0.85	1.88668707330119\\
0.86	1.76090812681997\\
0.87	1.63512914914206\\
0.88	1.50935014125743\\
0.89	1.38357110411352\\
0.9	1.2577920386198\\
0.91	1.13201294564792\\
0.92	1.00623382603382\\
0.93	0.880454680580378\\
0.94	0.754675510058189\\
0.95	0.628896315207517\\
0.96	0.503117096740613\\
0.97	0.37733785534052\\
0.98	0.25155859166589\\
0.99	0.125779306349216\\
};
\addlegendentry{$a = 0.0076$}

\addplot [color=orange,mark=square,mark repeat=4]
  table[row sep=crcr]{%
0.8	2.58040388949011\\
0.81	2.45138399647297\\
0.82	2.32236406526264\\
0.83	2.19334409714097\\
0.84	2.06432409333276\\
0.85	1.93530405500915\\
0.86	1.80628398329095\\
0.87	1.67726387925\\
0.88	1.54824374391364\\
0.89	1.41922357826607\\
0.9	1.29020338325063\\
0.91	1.16118315977127\\
0.92	1.03216290869604\\
0.93	0.903142630857712\\
0.94	0.774122327055693\\
0.95	0.645101998058206\\
0.96	0.516081644603528\\
0.97	0.387061267400612\\
0.98	0.258040867132359\\
0.99	0.129020444454964\\
};
\addlegendentry{$a = 0.0078$}

%

\end{axis}

\end{tikzpicture}%
}
			\caption{Global cost function $J$ against $a$ and $b$.}
			\label{fig:JAgainstExponentialDecreaseAndFloor}
		\end{figure}
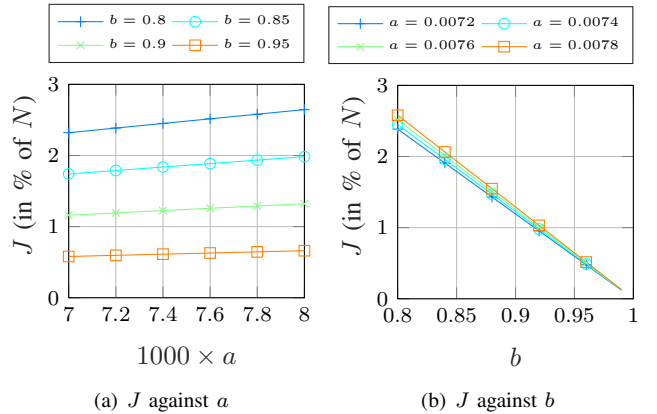

		Using the values of $a$ and $b$ from \figref{fig:ParametersabAgainstCarrierFrequency}, we show the behavior of $J$ against $a$ and $b$ in \figref{fig:JAgainstExponentialDecreaseAndFloor} assuming a PT-RS spacing of $\Delta=50$ in a signal of length $\N=4096$.
		
		\noindent
		$J$ exhibits a linear behavior with $a$ and $b$ which is convenient to predict its values. Similarly to the autocorrelation, see \ref{sec:EXPModel}, $J$ varies more with $b$ than with $a$. As a decrease in $b$ corresponds to an increase in $F_c$ that means an increase in the phase noise intensity, then, a decrease in $b$ involves a degradation of the Wiener filter performance while keeping the same values for $\Delta,\N$. This explains why $J$ is a decreasing function of $b$. 
		
		\begin{figure}[!h]
			\centering
			\vspace{1em}
			\input{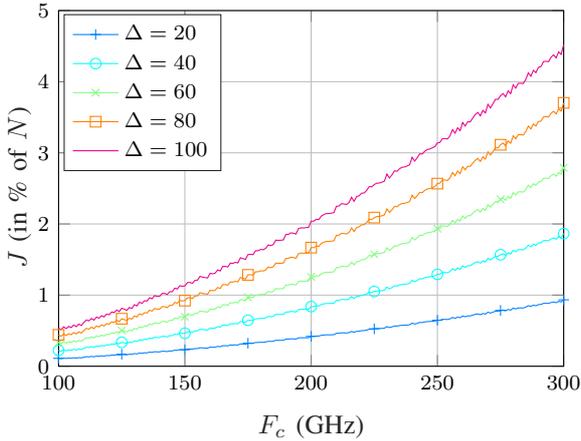}
			\caption{Global cost function against the carrier frequency $F_c$ for various values of $\Delta$.}
			\label{fig:JAgainstCarrierFrequency}
		\end{figure}
		
		Given that a value of the couple $(a,b)$ corresponds to a single value of $F_c$, we can show the evolution of $J$ with $F_c$ for various values of the PT-RS spacing $\Delta$, see \figref{fig:JAgainstCarrierFrequency}. This figure shows the maximum carrier frequency for a PT-RS spacing: when setting a maximum value for $J$, i.e. a maximum cost, we are indeed able to find the maximum carrier frequency value for which a given spacing leads to a Wiener filter with acceptable performance with respect to the said cost. As an example, when requiring that $J\leq 2\%$, the PT-RS spacing $\Delta=100$ becomes irrelevant when $F_c\geq 200$GHz.
		
	\subsection{Influence of $\Delta$}
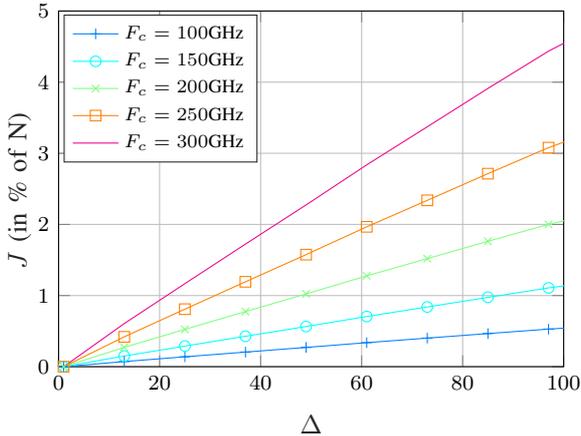
\begin{figure}[!h]
			\centering
%
%
\definecolor{mycolor1}{rgb}{0.00000,0.50000,1.00000}%
\definecolor{mycolor2}{rgb}{0.00000,1.00000,1.00000}%
\begin{tikzpicture}

\begin{axis}[%
	width=0.75\linewidth,
	height=0.2\textheight,
	font=\footnotesize,
	scale only axis,
	xmin=0,
	xmax=100,
	xlabel style={font=\color{white!15!black}},
	xlabel={$\Delta$},
	ymin=0,
	ymax=5,
	ytick={0,1,...,5},
	ylabel style={font=\color{white!15!black},anchor=south west,at={(0.15,0.25)}},
	ylabel={$J$ (in \% of $\N$)},
	axis background/.style={fill=white},
	xmajorgrids,
	ymajorgrids,
	legend style={at={(0.01,0.99)}, anchor=north west, font=\scriptsize, legend cell align=left, align=left, draw=white!15!black}
]
\addplot [color=mycolor1,mark=+]
  table[row sep=crcr]{%
1	2.02707641783206e-06\\
13	0.0721915932663464\\
25	0.13904144775202\\
37	0.205249418884701\\
49	0.270810587690651\\
61	0.337858204816732\\
73	0.401608301431544\\
85	0.465712263362994\\
97	0.528143836161177\\
109	0.580633707160072\\
};
\addlegendentry{$F_c =100$GHz}

\addplot [color=mycolor2,mark=o]
  table[row sep=crcr]{%
1	4.22453571014003e-06\\
13	0.150330664420917\\
25	0.289635072409689\\
37	0.427783335769572\\
49	0.564844407681475\\
61	0.705505968307718\\
73	0.839568563352011\\
85	0.974985499760783\\
97	1.1074200609417\\
109	1.21859248651378\\
};
\addlegendentry{$F_c =150$GHz}

\addplot [color=white!50!green,mark=x]
  table[row sep=crcr]{%
1	7.65414366332706e-06\\
13	0.272460242857586\\
25	0.524851413160932\\
37	0.774989299214657\\
49	1.0229292297363\\
61	1.27694661781814\\
73	1.51877074905684\\
85	1.76249933946547\\
97	2.00037750748958\\
109	2.20024045822937\\
};
\addlegendentry{$F_c =200$GHz}

\addplot [color=orange,mark=square]
  table[row sep=crcr]{%
1	1.17878853900244e-05\\
13	0.419631986485036\\
25	0.808335207050492\\
37	1.19353233692631\\
49	1.57529288073803\\
61	1.96631323043822\\
73	2.33850108053564\\
85	2.71349906630894\\
97	3.07938674149245\\
109	3.38683812667687\\
};
\addlegendentry{$F_c =250$GHz}

\addplot [color=magenta]
  table[row sep=crcr]{%
1	1.70219173423192e-05\\
13	0.606185308327067\\
25	1.1675336834695\\
37	1.72352369670679\\
49	2.27412740942595\\
61	2.83729509451847\\
73	3.37282074703318\\
85	3.91141620549373\\
97	4.43604840584075\\
109	4.87712922054349\\
};
\addlegendentry{$F_c =300$GHz}

\end{axis}

\end{tikzpicture}%
			\caption{Global cost function against the PT-RS spacing for various values of the carrier frequency $F_c$.}
			\label{fig:GlobalCostFunction}
		\end{figure}

		We observe in \figref{fig:GlobalCostFunction} the evolution of $J$ as a function of the PT-RS spacing $\Delta$ for various values of $F_c$. $J$ shows an affine behavior such that we can approximate it as follows:
		\begin{equation}
			J(\Delta) \approx \omega \Delta + \eta.
			\label{eq:GlobalCostAffine}
		\end{equation}
		
\begin{figure}[!h]
			\centering
			\input{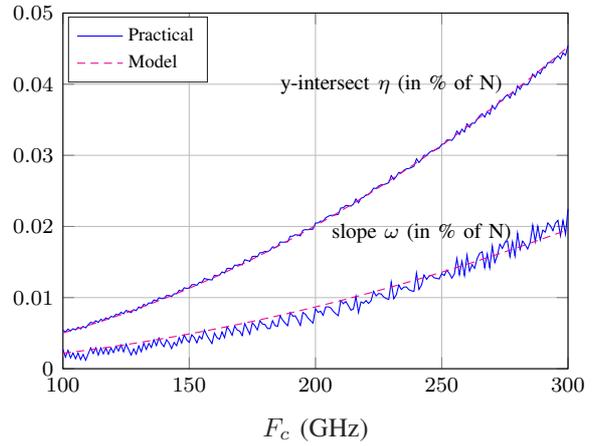}
			\caption{Slope and y-intersect of the linear approximation of $J(\Delta)$ against $F_c$ and an approximate model.}
			\label{fig:SlopeAndYintersectOfGlobalCostFunction}
		\end{figure}

		\noindent
		The slope $\omega$ and the y-intersect $\eta$ can be deduced from \eqref{eq:GlobalCostFunctionClosedForm}. However, at this stage, the affine reduction of \eqref{eq:GlobalCostFunctionClosedForm} is not tractable. Therefore, we obtain \eqref{eq:GlobalCostAffine} through a linear interpolation of $J$, see in \figref{fig:SlopeAndYintersectOfGlobalCostFunction} the evolution of $\omega$ and $\eta$ with the carrier frequency. The best model with respect to the mean square error is:
		\begin{eqnarray}
			\omega(F_c) & \approx & 5.03\cdot 10^{-25} F_c^2\;(\text{in \% of } \N), \\
			\eta(F_c) & \approx & 2.17\cdot 10^{-25} F_c^2\;(\text{in \% of } \N). \nonumber
		\end{eqnarray}
		
\section{PT-RS extraction\label{sec:PTRSExtraction}}
	In this section, we show how to extract $\lambda^*$ (or $\Delta^*$) that leads to the best performance of the Wiener filter in terms of the associated cost function defined in the previous section. This section exposes how to consider a performance constraint as well as an overhead constraint accompanied with an application example.
	
	\subsection{Constraints\label{sec:Constraints}}
		To ensure a minimum data rate that corresponds to a reasonable PT-RS overhead, we consider a minimum PT-RS spacing $\Delta_0$. As $J$ is an increasing function of $\Delta$, setting $\Delta=\Delta_0$ surely performs the best. Yet, the Wiener filter may have acceptable performance even when enlarging the spacing up to a maximum value $\Delta_\text{PF}$. In other words, we assert that while $\Delta_0\leq\Delta\leq\Delta_\text{PF}$, the Wiener filter well performs with $\Delta$. In the next paragraph, we show how to determine the value $\Delta_\text{PF}$ based on the result from the previous section.
	
		As the PT-RS distribution is uniform, we can assume that the number of PT-RS for a given spacing $\Delta$ is $\NP = \lceil \frac{\N}{\Delta} \rceil$. The PT-RS overhead being the ratio between $\NP$ and $\N$, we deduce that it is closed to $\frac{1}{\Delta}$. 
	\subsection{Example of use}
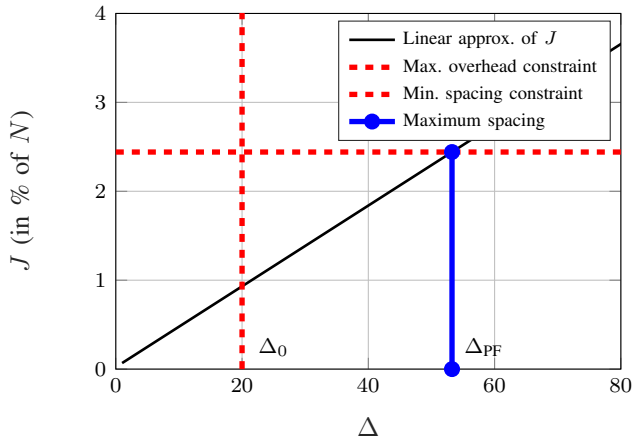
\begin{figure}[!h]
			\centering
%
%
\begin{tikzpicture}

\begin{axis}[%
	width=0.75\linewidth,
	height=0.2\textheight,
	font=\footnotesize,
	scale only axis,
	xmin=0,
	xmax=80,
	xlabel style={font=\color{white!15!black}},
	xlabel={$\Delta$},
	ymin=0,
	ymax=4,
	ytick={0,1,2,3,4},
	ylabel style={font=\color{white!15!black}},
	ylabel={$J$ (in \% of $N$)},
	axis background/.style={fill=white},
	xmajorgrids,
	ymajorgrids,
	legend style={legend cell align=left, align=left, draw=white!15!black, font=\scriptsize}
]
\addplot [color=black, line width=1.0pt]
  table[row sep=crcr]{%
1	0.0678772361524285\\
13	0.612928890191361\\
25	1.15798054423029\\
37	1.70303219826923\\
49	2.24808385230816\\
61	2.79313550634709\\
73	3.33818716038603\\
85	3.88323881442496\\
97	4.42829046846389\\
109	4.97334212250282\\
};
\addlegendentry{Linear approx. of $J$}

\addplot [color=red, dashed, line width=2.0pt]
  table[row sep=crcr]{%
20	0\\
20	4.8828125\\
};
\addlegendentry{Max. overhead constraint}

\addplot [color=red, dashed, line width=2.0pt]
  table[row sep=crcr]{%
0	2.44140625\\
120	2.44140625\\
};
\addlegendentry{Min. spacing constraint}

\addplot [color=blue, line width=2.0pt, mark size=2.0pt, mark=*, mark options={solid, fill=blue, blue}]
  table[row sep=crcr]{%
53.2562365513643	0\\
53.2562365513643	2.44140625\\
};
\addlegendentry{Maximum spacing}

\node at (axis cs:25,0.25) {$\Delta_0$};
\node at (axis cs:58.256,0.25) {$\Delta_\text{PF}$};
\end{axis}

\end{tikzpicture}%
			\caption{Example of the optimal pilot pattern extraction at $F_c=300$GHz.}
			\label{fig:Example}
		\end{figure}

		We consider here that an experimenter requires the communication chain to perform at $300$GHz such that the Wiener filter performs such that the cost $J$ remains below $MaxCost=2.5\%$. Moreover, the experimenter requires a minimum throughput such that the PT-RS spacing should not decrease below $\Delta_0 = 20$ assuming signals of length $\N=4096$. By choosing $\Delta=\Delta_0$ we obtain a PT-RS overhead of $5\%$. From the previous approximations, we can set:
		\begin{eqnarray}
			\omega(F_c=300\text{GHz}) & = & 0.0453\;(\text{in \% of $\N$}),\\
			\eta(F_c=300\text{GHz}) & = & 0.0195\;(\text{in \% of $\N$}).
		\end{eqnarray}
		We observe that $J(\Delta_0) \approx 0.9248\%$ which is below $MaxCost$. Therefore, we can propose a PT-RS spacing larger than $\Delta_0$ to fulfill the cost constraint while ensuring an acceptable expected spectral efficiency. By inverting the affine approximation, we obtain the maximum spacing $\Delta_\text{PF}$:
		\begin{equation}
			\Delta_\text{PF} = \left\lfloor \frac{MaxCost-\eta(F_c=300\text{GHz})}{\omega(F_c=300\text{GHz})} \right\rfloor = 54.
		\end{equation}
		With this spacing, the resulting PT-RS overhead is reduced to $1.85\%$ which leaves a significant space for the data symbols. This example is shown in \figref{fig:Example}.

\section*{Conclusion\label{sec:Conclusion}}
	In this paper, we presented a method to obtain the spacing between PT-RS positions that allows acceptable performance of the Wiener filter, used to track the phase noise, while ensuring a sufficiently small pilot overhead. We considered an analytical derivation of the cost function whose behavior is monotonic, leading to a straightforward selection of the spacing given the said constraints on the overhead and the performance. The derivation relied on an exponential approximation of the autocorrelation of the 3GPP phase noise model. In future works, one could target an enhancement of the said approximation to better fit with the model for very high values of the carrier frequencies toward THz.

\bibliographystyle{IEEEtran}
\bibliography{IEEEabrv,biblio}

\end{document}